**Title:** AI for NONMEM Coding in Pharmacometrics Research and Education: Shortcut or Pitfall?

**Authors:** Wenhao Zheng[1], Wanbing Wang[2,3], Carl M.J. Kirkpatrick[4], Cornelia B. Landersdorfer[4], Huaxiu Yao[1,5], Jiawei Zhou[2,6]

**Affiliations:**

[1]Department of Computer Science, University of North Carolina at Chapel Hill, Chapel Hill, North Carolina, USA

[2]Dvision of Pharmacotherapy and Experimental Therapeutics, Eshelman School of Pharmacy, University of North Carolina at Chapel Hill, Chapel Hill, North Carolina, USA

[3]Department of Biostatistics, University of North Carolina at Chapel Hill, Chapel Hill, North Carolina, USA

[4]Monash Institute of Pharmaceutical Sciences, Monash University, Parkville, VIC, Australia

[5]School of Data Science and Society, University of North Carolina at Chapel Hill, Chapel Hill, North Carolina, USA

[6]Lineberger Comprehensive Cancer Center, School of Medicine, University of North Carolina at Chapel Hill, Chapel Hill, North Carolina, USA

**Corresponding Author\*:**

Jiawei Zhou, University of North Carolina at Chapel Hill, 2320 Kerr Hall, Chapel Hill, NC, 27599, USA

Email Address: zhoujw@unc.edu

**Conflicts of Interests:**

The authors report no conflicts of interest.

**Abstract**

Artificial intelligence (AI) is increasingly being explored as a tool to support pharmacometric modeling, particularly in addressing the coding challenges associated with NONMEM. In this study, we evaluated the ability of seven AI agents to generate NONMEM codes across 13 pharmacometrics tasks, including a range of population pharmacokinetic (PK) and pharmacodynamic (PD) models. We further developed a standardized scoring rubric to assess code accuracy and created an optimized prompt to improve AI agent performance. Our results showed that the OpenAI o1 and gpt-4.1 models achieved the best performance, both generating codes with great accuracy for all tasks when using our optimized prompt. Overall, AI agents performed well in writing basic NONMEM model structures, providing a useful foundation for pharmacometrics model coding. However, user review and refinement remain essential, especially for complex models with special dataset alignment or advanced coding techniques. We also discussed the applications of AI in pharmacometrics education, particularly strategies to prevent over-reliance on AI for coding. This work provides a benchmark for current AI agents performance in NONMEM coding and introduces a practical prompt that can facilitate more accurate and efficient use of AI in pharmacometrics research and education.

**Introduction**

With the rapid advancement of artificial intelligence (AI), AI agents (the AI software that could chat like a person) have shown great potential in helping with complex tasks such as writing program codes. (1) These AI tools are designed to understand and respond in plain language, making them increasingly useful to improve coding efficiency. In the area of pharmacometrics, which involves coding mathematical models to understand how drugs behave in the body, researchers are beginning to explore how AI can help with model development. (2, 3)

One of the most widely used tools in pharmacometrics is NONMEM, a specialized software for population pharmacokinetic/pharmacodynamic (PK/PD) analysis using nonlinear mixed-effects modeling. (4) However, writing NONMEM code poses challenges for individuals who lack prior programming experience or a solid understanding of population PK/PD modeling approaches. Nowadays, pharmacometricians are exploring ways to incorporate AI agents into daily drug development activities and educational settings to ease coding demands. Recent studies have evaluated the performance of AI agents such as gpt-4.0 and Gemini to generate NONMEM code, and consistently highlight a need for improved accuracy and expert oversight in using AI for NONMEM coding. (5-7) These studies provide valuable insights into the potential of AI-generated NONMEM scripts, and highlight the need for further evaluation and refinement to improve the accuracy and reliability of AI-assisted NONMEM coding.

Building on this foundation, we developed a standardized scoring rubric to systematically evaluate AI agents' performance in NONMEM coding tasks. We also created an optimized prompt that substantially enhances the accuracy and usability of AI-generated code. This approach is designed to support pharmacometricians and clinical pharmacologists in tasks such as population PK/PD modeling, trial simulation, and covariate analysis, enabling real-world applications in drug development, regulatory submissions, and dose optimization. We also discussed the broader implications of integrating AI tools into pharmacometrics education, highlighting both its potential benefits in enhancing NONMEM training and accessibility, as well as the associated risks, including overreliance on automated tools and the need for expert oversight.

**Key Questions**

This study aims to address several key questions by evaluating and comparing seven AI agent models: How accurate are current AI agents in generating NONMEM codes? Which AI agent demonstrates the best performance? Can the accuracy of AI-generated NONMEM codes be improved, and if so, how?

**Analysis Plan**

To evaluate the performance of current AI in writing NONMEM codes, we created 13 coding tasks for the AI agents to complete. These tasks are listed in **Table 1** and include building various types of pharmacokinetic (PK) and pharmacodynamic (PD) models—such as one-, two-, and three-compartment PK models, direct and indirect response PD models, and more complex models such as target-mediated drug disposition (TMDD) models. We evaluated the accuracy of completing these tasks using seven AI agents: gpt-4.1-mini, gpt-4.1-nano, gpt-4.1, gpt-4o-mini, gpt-4o, o1, and o3-mini. These models were selected to capture a range of capabilities across different AI model families, including both high-performance and lightweight variants.(8) This selection allowed us to assess performance variability and generalizability across commonly available AI tools. All of these AI agents were developed by OpenAI and are available under a commercial license, but they differ in how they are built (architecture), how large and complex they are (size), and how well they can understand and generate responses (capacity). (9) We used each AI agent to generate NONMEM codes for all 13 tasks.

We then developed a standard scoring rubric to evaluate task performance. The rubric was developed iteratively using several preliminary test tasks and was informed by common AI-generated NONMEM coding errors, such as incorrect structural model syntax, misalignment with dataset variables, and misuse of NONMEM subroutines. It was then refined and validated through expert consensus to ensure it captured clinically and technically meaningful distinctions in code accuracy.

To explore ways to improve AI performance, we transformed the predefined rubric into an optimized prompt, embedding its criteria directly into the prompt structure. A prompt is the instruction or question you give to the AI to guide how it responds. The optimized prompt, provided in the **Supplementary Materials**, was used to re-run the same 13 tasks. We compared the results to determine whether the optimized prompt could improve the accuracy of AI-generated NONMEM codes.

**Results**

The final scoring rubric included three levels of accuracy based on whether the codes can correctly specify model components, define parameter relationships, and use appropriate subroutines. (10, 11) As shown in **Table 2**, points were assigned for each criterion met in the rubric, with a maximum score of 6 indicating that the code fully meets all requirements with high accuracy.

We evaluated and compared the performance of each response using the standardized scoring rubric (**Figure 1**). Among all the AI agents, o1 demonstrated the best overall performance, with a mean score of 5 (interquartile range [IQR] 2.5–6) using the original prompt and a perfect score of 6.0 in all 13 tasks using the optimized prompt. The gpt-4.1 AI agent ranked second, scoring 5 (IQR 2.5–5) with the original prompt. It also achieved a perfect score of 6.0 in all 13 tasks with the optimized prompt. Notably, both o1 and gpt-4.1 were able to generate NONMEM codes for all 13 tasks with great accuracy when guided by the optimized prompt, highlighting their strong coding capabilities. Responses generated by all seven AI agents, as well as their responses using the optimized prompt, are provided in the **Supplementary Materials**.

We also analyzed task-specific improvements with the optimized prompt and found that it enhanced coding accuracy across all AI agents, with a greater improvement seen in more complex models such as indirect response and PK models with absorption lag time (**Figure 2**).

Supplementary Materials can be found https://github.com/zhoujw14/AI_PMx.

**Impact Assessment**

In this study, we developed a standardized scoring rubric to evaluate the accuracy of AI-generated NONMEM code and compared the performance of seven AI agents developed by OpenAI. Among these models, o1 demonstrated the highest overall performance, followed closely by gpt-4.1, both of which showed strong capabilities in generating pharmacometrics model codes. To further enhance performance, we also designed optimized prompts that significantly improved code accuracy. The optimized prompt can be used in future NONMEM code generation tasks to support both research and education in pharmacometrics.

Our findings suggest that AI agents are increasingly capable of generating accurate NONMEM code for basic model structures and parameter specifications. Based on our evaluations, the AI-generated scripts can serve as a solid starting point for model development. However, user review and editing remain essential. In particular, the AI-generated codes may still require refinement to ensure alignment with specific datasets, study designs, or modeling objectives. For more complex modeling tasks such as indirect response models or other PK/PD models, current AI agents show limitations. For example current AI agents may fail to implement differential equations correctly or misapply model compartments. They may also struggle with proper initialization of model parameters or aligning dosing/event records in the dataset, leading to incorrect model execution. This highlights the need for future AI agents trained specifically on NONMEM guidelines (12) and pharmacometric coding conventions to achieve greater accuracy and reliability.

In the context of pharmacometrics education, the growing capabilities of AI tools offer an opportunity to lower the entry barrier to modeling by reducing the emphasis on technical coding skills. The focus on pharmacometrics education should shift from "technical coding skills" to

"quantitative thinking", with a focus on the ability to understand clinical questions, develop models that are fit for purpose, and effectively communicate modeling results. AI agents can serve as valuable tools for learning, allowing students to generate initial model drafts, test different approaches, and receive real-time feedback. To maximize educational benefit, students should also be trained to craft effective prompts and critically assess the quality and correctness of AI-generated codes. However, there are potential risks associated with overreliance on AI tools. For example, users may unknowingly accept plausible looking but technically incorrect codes, especially in complex models involving data pre-processing or advanced estimation methods. This highlights the need for expert oversight and continued education to ensure appropriate interpretation and use of AI-generated output. Future development of AI agents for pharmacometrics education should emphasize critical thinking and reasoning capabilities, rather than focusing solely on generating error-free code. Ideally, AI agents should be designed to support model reasoning and literature integration, serving as educational tools that enhance understanding rather than replace it.

Another promising application of AI in education is its potential use as an automated code screening tool. AI agents could help identify major errors or omissions in NONMEM code, providing students with immediate feedback before expert review. While this application holds potential, the variability and flexibility of NONMEM syntax and modeling strategies mean that further refinement of AI agents is needed to ensure consistent and reliable code evaluation.

**Conclusion**

In summary, we evaluated the performance of seven different AI agents in generating NONMEM codes based on a standardized scoring rubric and introduced an optimized prompt to enhance their accuracy. Our study not only benchmarks the current capabilities of AI agents in pharmacometric modeling but also provides practical prompts for AI agents to support the broader integration of AI in pharmacometrics research and education. Continued development and domain-specific training of AI agents will be essential to fully realize their potential in our field.

## Acknowledgements

This work is funded by a PharmAlliance Early Career Research Award and University of North Carolina at Chapel Hill.

**Table 1. NONMEM coding tasks assigned to the AI agent.**

| ID | Coding Task |
|---|---|
| 1 | Write a one-compartment, first-order absorption, linear clearance NONMEM model |
| 2 | Write a two-compartment, first-order absorption, linear clearance NONMEM model |
| 3 | Write a one-compartment, first-order absorption, nonlinear clearance NONMEM model |
| 4 | Write a three-compartment, first-order absorption, linear clearance NONMEM model |
| 5 | Write a target-mediated drug disposition NONMEM model |
| 6 | Write a absorption with lag time, one-compartment linear PK NONMEM model |
| 7 | Write a one-compartment, i.v. infusion, linear clearance NONMEM model |
| 8 | Write a one-compartment, transit absorption, linear clearance NONMEM model |
| 9 | Write a exponential tumor growth model in NONMEM |
| 10 | Write a two-compartment, transit absorption, linear clearance NONMEM model |
| 11 | Write a direct exposure-response NONMEM model |
| 12 | Write an indirect response model, with stimulatory effect on Kin, NONMEM model |
| 13 | Write an indirect response model with placebo effects, with stimulatory effect on Kin, NONMEM model |

**Table 2. Scoring rubric for AI-written NONMEM codes.**

| Points | Rubric |
|---|---|
| 1 | **Level 1. Include the essential NONMEM control stream blocks** $PROB, $INPUT, $DATA, $SUBROUTINES, $PK, $ERROR, $THETA, $OMEGA, $SIGMA, $ESTIMATION (or the shorthand $EST), $COVARIANCE (or $COV), and $TABLE. |
| 2 | **Level 2: Ensure the correct relationship between THETA and ETA is included** Ensure the correct relationship between THETA and ETA is included. For each ETA(n), there should be a corresponding line in the $PK block. |
| 3 | **Level 3: $SUBROUTINE, $DES, $PK and parameters setting follow the NONMEM 7.4 (PREDPP guide VI). (11)** Table 1 from reference (10) were used as the input in this rubric. |

**Figure Legends**

**Figure 1. Summary of AI Agent Performance.** NONMEM codes generated by different AI agents for 13 tasks were evaluated using a predefined scoring rubric. Each boxplot displays the median score (vertical line within the box) and the interquartile range. The whiskers represent the minimum and maximum scores across the 13 tasks. The top-performing AI agents are indicated with # symbol. IQR, interquartile range.

**Figure 2. AI agent performance improved with optimal algorithm in writing NONMEM codes.** The score of each question was compared between original AI agent and AI agent optimized using rubric prompt.

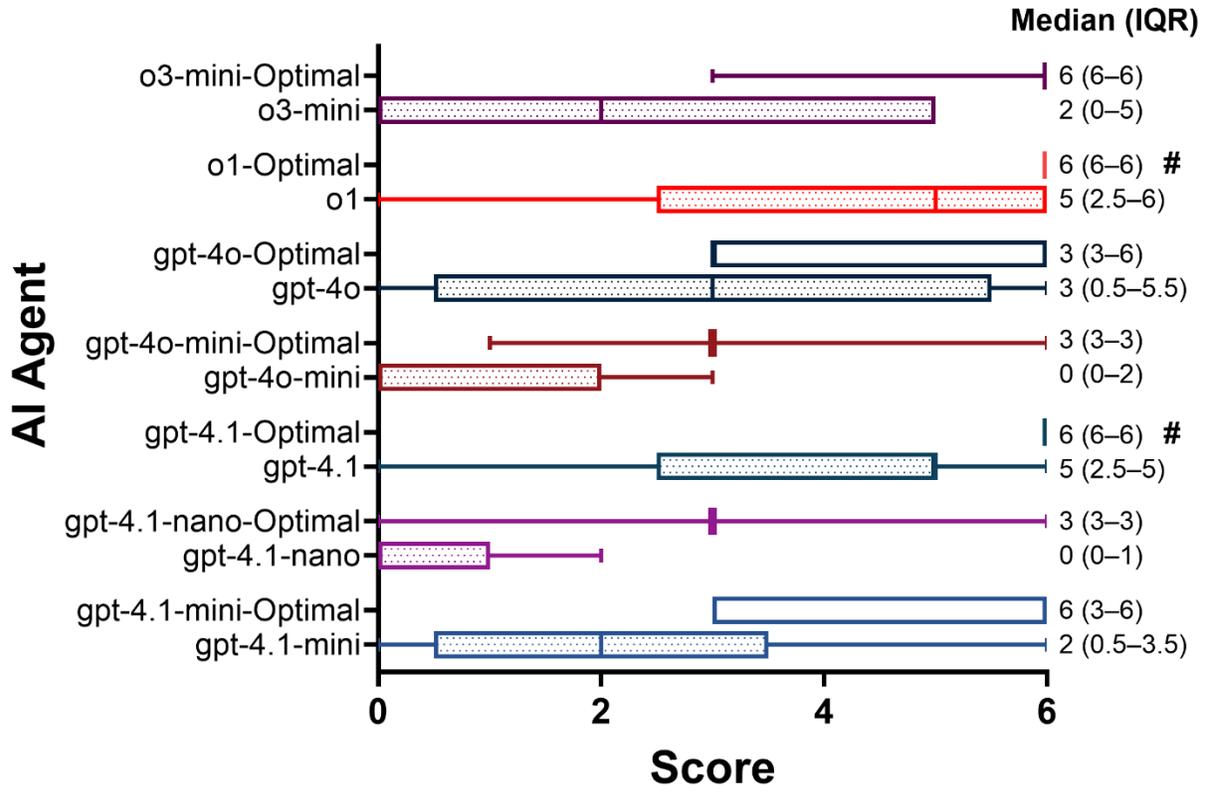

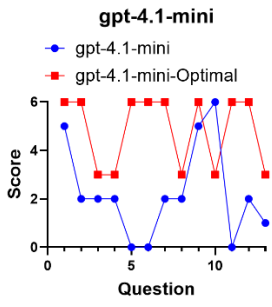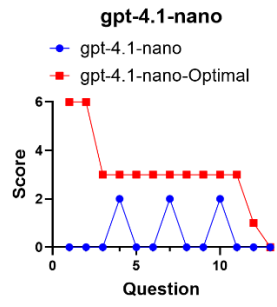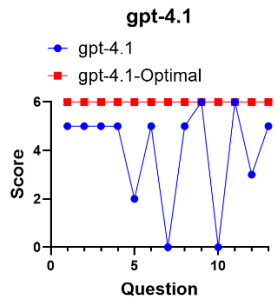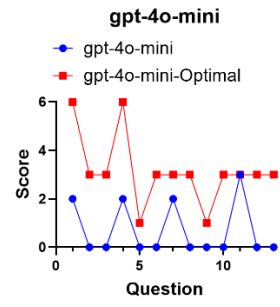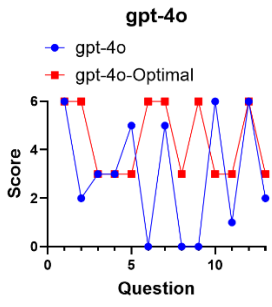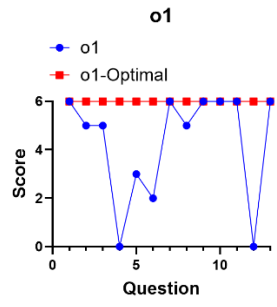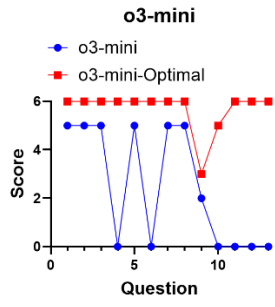